# Machine learning classification of power converter control mode

Rabah Ouali  Jean-Yves Dieulot  Pascal Yim  Xavier Guillaud
Frédéric Colas  Yang Wu  Heng Wu

January 19, 2024

Abstract

To ensure the proper functioning of the current and future electrical grid, it is necessary for Transmission System Operators (TSOs) to verify that energy providers comply with the grid code and specifications provided by TSOs. A lot of energy production are conntected to the grid through a power electronic inverter. Grid Forming (GFM) and Grid Following (GFL) are the two types of operating modes used to control power electronic converters. The choice of control mode by TSOs to avoid impacting the stability of the grid is crucial, as is the commitment to these choices by energy suppliers. This article proposes a comparison between commonplace machine learning algorithms for converter control mode classification: GFL or GFM. The classification is based on frequency-domain admittance obtained by external measurement methods. Most algorithms are able to classify accurately when the control structure belongs to the training data, but they fail to classify modified control structures with the exception of the random forest algorithm.

## 0.1 Introduction

Due to increasing efforts to mitigate the effects of climate change, the electricity production sector is undergoing a decarbonization phase through the integration of carbon-neutral energy sources. The majority of these sources, such as photovoltaic and wind energy, are connected to the electrical grid using power electronic converters. This presents numerous challenges for the grid, including the transition from a network dominated by synchronous generators to one with a high penetration of distributed converters in a heterogeneous manner [1]. The behavior of a converter connected to the grid is closely tied to its control mode. It belongs to two main classes: Grid Forming (GFM) and Grid Following (GFL) [2]. Nowadays, the GFL control mode which makes the converter behave as a current source is widespread. However, The growing adoption of this converter control method in the production mix results in reduced inertia and heightened sensitivity of the converter to weak grids [3]. Another kind of control mode, called GFM allows the converter to behave as a voltage source which is able to offer voltage support to the grid. However, a high level of GFL or GFM in the grid can bring stability issues [4], and it is crucial for the Transmission System Operator (TSO) to choose, at a given location, in which mode a converter will operate. The TSO has to check whether the configuration of the converters are set accordingly. However, TSOs have a limited access to their structures and control parameters because converters are owned by private suppliers and are protected by intellectual property rights. Therefore, TSOs have to find out the dynamical behavior of the converters using external measurements. A modeling method based on measurements at the Point of Common Coupling (PCC), using impedance (IM) or admittance (AM) models, has been proposed to analyze the behavior of converters connected to the grid without accessing the control and considering them as black-boxes [5]. In [6], Gong et al. presented an external measurement method to identify the impedance model of a converter connected to the grid as a black-box. This method was validated by comparing its results with the analytical model obtained by a white-box approach. The measured impedance model was also used to analyze the stability of a single converter and the interaction between different converters connected to the same network [5]. However, the impedance results depend on the operating point and the value of the control parameters. Hence, to assess the performance of a converter across various operating points, it is essential to replicate the measurement procedure for each operating point. Qiu et al. [7] have developed a method to find the admittance model at multiple operating points with only measurements at one operating point and the use of an optimization method. This procedure has proven to be effective but requires a considerable amount of computation time and is not able to identify the used mode of control algorithms or the control parameters. In [8], a method for identifying the control parameters of a converter operating in GFL mode, based on an optimization model is proposed. This method is valid only at specific operating points and for specific control structures. To conclude, while identification methods have been extensively examined in the literature, there has been no exploration of identifying the type of black-box control mode based on external measurements at the PCC. The objective of this article is to evaluate the capability of machine learning (ML) algorithms to identify the control mode used by a converter connected to the grid based on the measured black-box admittance model and commonplace ML algorithms. The rest of the paper is organized as follows: Section 2 presents the control mode and structures of the power converter. Section 3 presents the measurement of frequency admittance model. Section 4 presents dataset generation and control parameter variation. Section 5 presents the results of machine learning classification.

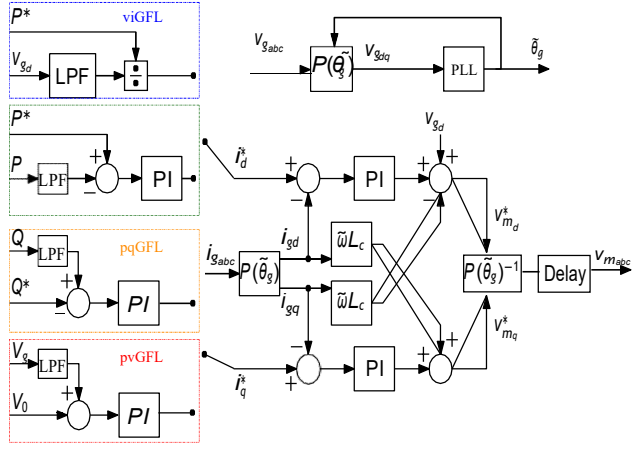

Figure 1: pqGFL and pvGFL control structures.

## 0.2 Power converter control modes and structures

The power converters in GFL mode are controlled as current sources. GFL operates within an existing grid that sets the voltage and frequency. A Phase-Locked Loop (PLL) is used for synchronization, and the output currents of the converters are controlled through different control loops [9]. Several control structures are proposed to control the output currents, with Active and Reactive power controde (pqGFL), active and voltage control mode (pvGFL), and viGFL being the most commonly used, as illustrated in Fig. 1. For both pqGFL and pvGFL structures, the reference for the direct current $I_d^*$ is generated by an active power control loop. However, for viGFL, the active power loop is replaced by a simple division, where the active power reference is divided by the direct component of the voltage measured at the PCC and filtered through a first-order low-pass filter. However, the reference for the quadrature current $I_q^*$ is generated in both pqGFL and viGFL using a reactive power control loop, while for pvGFL, it is generated based on a voltage control loop at the PCC. The two current references for all three structures are then provided to a current control loop that regulates the output current of the converter.

For the GFM mode, the objective is to control the converter as a voltage source. To achieve this functionality, the frequency and voltage output of the converter are controlled with feedback loops. The frequency output of the converter is typically controlled with the help of an active power loop. This power loop can be achieved by several control structures, including those aimed at emulating the operation of a synchronous machine [9]. For voltage control, multiple control structures are also proposed, the most widely used being: Voltage Controlled GFM (vcGFM) and Current Controlled GFM (ccGFM). For the vcGFM structure, the modulated voltage is directly defined, with an additional damping resistor, while ccGFM relies on a quasi-static model to generate the current reference, which is then regulated by a current control loop as shown Fig. 2. All the aforementioned control structures will be considered in this article.

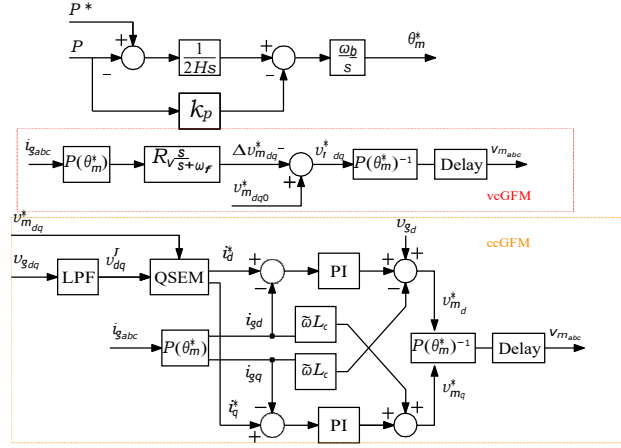

Figure 2: ccGFM and vcGFM control stuctures.

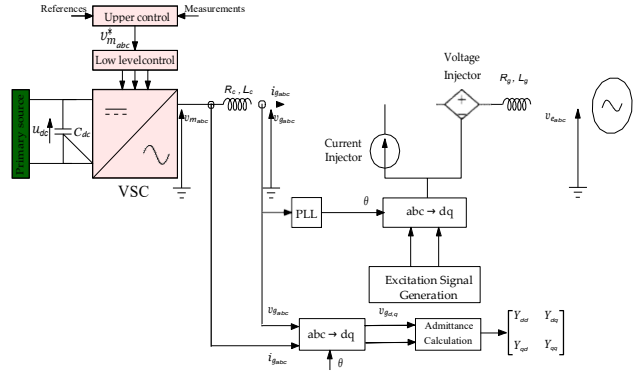

Figure 3: System diagram of the admittance measurement

## 0.3 Measurement of admittance model in the frequency domain of converters

Several methods have been proposed to measure the frequency-domain admittance model of the converter at the PCC, including shunt, series, and hybrid methods. A hybrid excitation configuration is presented in Fig. 3 to measure the converter's AM in the d-q reference frame. The basic principle of this method involves injecting a disturbance at the PCC and then calculating the AM based on the response of the voltage $V_{dq}$ and the current $I_{dq}$ measured at the PCC. The detailed procedure for measuring the AM is presented in Fig. 4.

To verify the accuracy of the external measurement method, a configuration for AM measurement on a vcGFM converter, as illustrated in Fig. 3, was simulated using Matlab/Simulink. The measurement frequency range extends from 1 Hz to 800 KHz. Tab. 1 presents the parameters of the tested converter.

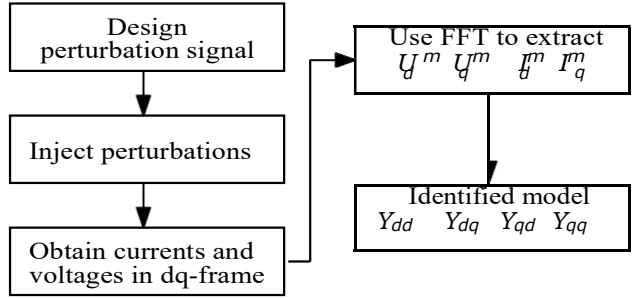

Figure 4: Flowchart of the admittance measurement.

Table 1: Circuit parameters

| Control structures | Parameter | Value |
|---|---|---|
| VSC | $S_{nom}$, $S_b$ | 1.044 GVA |
| | $U_{nom}$, $U_b$ | 400 KV |
| | $P_{nom}$ | 1 GW |
| | $Q_{max}$ | 300 MVar |
| Filter | $L_c$ | 0.15 pu |
| | $R_c$ | 0.005 pu |
| Excitation signal | $PRBS_d$: Order, $F_{max}$ | 8, 1000 Hz |
| | $PRBS_q$: Order, $F_{max}$ | 7, 1020 Hz |
| Grid | $L_c$ | 0.5 pu |
| | $R_c$ | 0.05 pu |

The results of the d,q admittance measurement for the converter using a Pseudo-Random Binary Sequence (PRBS) perturbation signal are compared to the analytical model of the admittance developed from small-signal system equations, serving as a reference for the measurement results.

Fig. 5 presents the measurement results and the small-signal analytical model for the $Y_{dd}$ admittance. The measurement results closely match the analytical model. This demonstrates that the external method can be used for system admittance measurement without the need to access the control structure and control parameters.

### 0.3.1 Influence of control parameters on AM

The frequency range from 1 Hz to 1 kHz, within which the converter control operates, is generally found form physical constants or grid code requirements. Fig. 6 illustrates three gain margins for AM in the dd-axis: two for vcGFM and pqGFL control stuctures, and a third for the passive L-filter at the converter's output. In both stuctures, the converter's admittances converge at the base frequency to the same value, depending on the operating point.

At higher frequencies, the admittances always converge towards the passive output filter's admittance. However, the AM within the range between 1 Hz and 1 kHz strongly depends on the behavior of the control mode and its associated parameters. This analysis highlights the significant influence of control algorithms on the converter's admittance model within the frequency range and precisely defines the interval of control influence on AM.

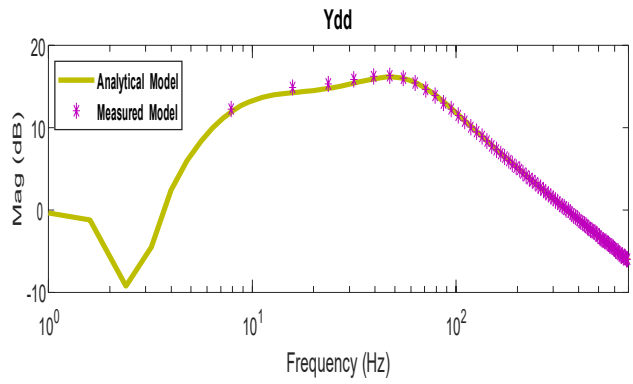

Figure 5: Analytical and measured models.

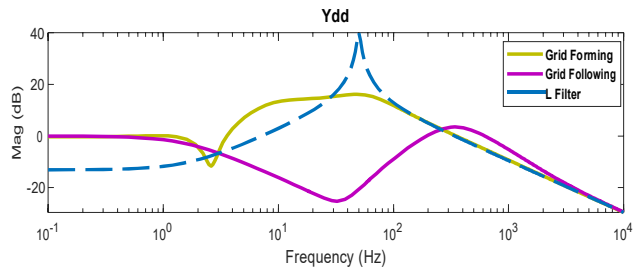

Figure 6: The variation of the converter's admittance with respect to frequency.

Table 2: Grid Following parameters

| Control structures | Parameter | Value |
|---|---|---|
| Outer loop | $\omega_p$ | [6 38] rad/s |
|  | $\omega_q$ | [6 38] rad/s |
|  | $\omega_v$ | [3 15] rad/s |
| PLL | $\omega_n$ | [50 1500] rad/s |
|  | $\xi_n$ | 1 |
| Current loop | $\omega_{cc}$ | [1200 3000] rad/s |
| Delay | d | [50 200] µs |

## 0.4 Dataset generation

After validating the external measurement method for the AM and defining the model's range of variations based on changes in control parameters, datasets are generated using simulation models developed in MATLAB Simulink. The creation of these datasets involves adjusting control parameters within specified intervals while adhering to physical constraints and grid code specifications. Typically, active power fluctuates between -1 pu and 1 pu during normal operating conditions, with participation in reactive power ranging from -0.4 pu to 0.4 pu. Voltage variation in the grid is generally limited to 10% of the nominal voltage, following grid code standards. In this study, the grid impedance parameters are held constant because the mixed-method-based measured model is independent of the grid impedance's influence. The grid parameter is defined with a Short-Circuit Ratio (SCR) of 15. Similarly, the passive converter filter parameters also remain constant because, as mentioned earlier, the value of the filter inductance does not significantly affect the measurement range. The values of grid impedance and converter are specified in Tab. 1.

In Tab. 2, the intervals of variation for the control parameters of three GFL structures, pqGFL, pvGFL and viGFL, are presented. The parameters $\omega_p$, $\omega_q$, $\omega_v$, and $\omega_{cc}$ represent the bandwidths for active power control, reactive power control, voltage control, and current control, respectively. The gains of the PLL controller are calculated based on the natural frequency $\omega_n$ and damping $\xi_n$. Furthermore, the delay associated with low-level control is considered due to its influence on the AM.

Tab. 3, presents the intervals of variation for the control parameters of two GFM structures, ccGFM and vcGFM. The parameters $\omega_{cc}$ and $\omega_{LPF}$ represent the bandwidths of current control and the quasi-static model, respectively. Additionally, $H$ represents inertia, and $\xi$ represents the damping of the active power loop. The delay associated with low-level control is also considered and varies within the same range as GFL.

Fig. 7 shows AM in dd axis of both structures ccGFM and pqGFL. The variation in control parameters of both modes creates overlaps between the models of these modes in certain configurations. These overlaps sometimes occur on both the dd and qq axes, or only on a single axis. Due to these overlaps, classifying the two control modes is not easy without using ML.

After generating multiple simulations with different control parameters for both control modes and the operating power, the results of each simulation are saved, as shown in Fig. 8. The magnitude and phase of both admittances $Y_{dd}$ and $Y_{qq}$ are stored, the coupling axes $Y_{qd}$ and $Y_{dq}$ are not considered in this study.

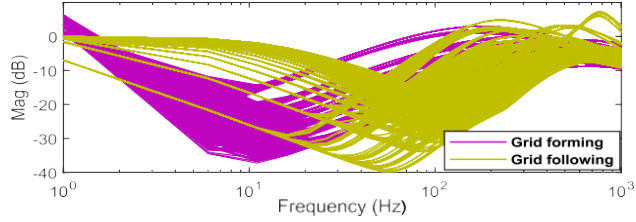

Figure 7: AM of ccGFM and pqGFL in the range of parameters.

Table 3: Grid Forming parameters

| Control structures | Parameter | Value |
|---|---|---|
| Power loop | H | [0.5 5] s |
|  | ξ | [0.7 4] |
| TVR | $\omega_f$ | 60 rad/s |
|  | R | 0.09 |
| QSEM | $\omega_{LPF}$ | 83 rad/s |
| Current loop | $\omega_{cc}$ | [1200 3000] rad/s |
| Delay | d | [50 200] μs |

The control mode of each measurement is saved in the last column "mode", is the output of the training set for learning algorithms.

## 0.5 Classification with ML

As mentioned earlier, the objective of this article is to classify the control modes of power converters using ML techniques. The target class in this study is known, and the goal is classification, which means that supervised ML algorithms are employed [10]. Decision Trees (DT), Random Forests (RF), Logistic Regression (LR), Support Sector Machines (SVM), k-Nearest Neighbors (KNN), Xgboost (XGB), and the Naive Bayes Classifier (NBC) are used in this study [10]. Furthermore, the classification results are compared for evaluation.

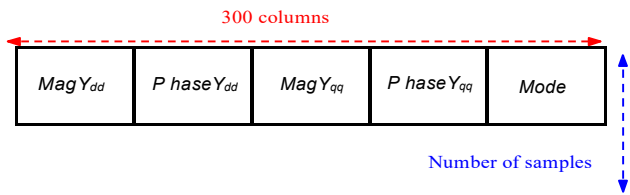

Figure 8: Dataset structure.

### 0.5.1 Steps in the classification process

The classification procedure is quite usual and consists of four steps:

1. Clean the data, handling missing values, and normalize it, especially for certain algorithms like logistic regression and support vector machines (SVM).

2. Split the data into two parts, one for training and the other for testing. In this article, 80% of the data is allocated for training, and 20% for testing.

3. Configure the model for each ML algorithm by defining its parameters, such as the number of trees in the Random Forest algorithm.

4. Train the ML model with the training data and validate it using the test data.

### 0.5.2 Classification Results

The total number of samples is 5000 for each control mode, divided among the four control structures: cvGFM and ccGFM for GFM mode, and pvGFL and pqGFL for GFL mode. The dataset of viGFL structure are not utilized in the classification data. Table 4 presents the parameters of each developed ML model and the classification accuracy. The classification results for each ML algorithm represent the average cross-validation score over 100 runs. Various ML algorithms have successfully classified the two control modes. The classification results are satisfactory due to the distinct control structures and parameters of the two modes. Additionally, the admittance models of the two modes are unique, facilitating their straightforward classification.

Table 4: Model parameters and Classification results

| ML algorithm | Parameters | Accuracy |
|:---:|:---:|:---:|
| LR | max_iter=100 | 96% |
| RF | n_estimators=100 | 98% |
| DT | max_depth=6 | 92% |
| NBC | By default | 92% |
| XGB | kernel=linear | 95% |
| SVM | kernel=linear | 96% |
| KNN | n_neighbors=20 | 96% |

In general, control modes (GFL, GFM) can be implemented with various control structures, as detailed in section 2. Changes in the control structure lead to alterations in the admittance model, and even minor adjustments in the control structure can result in variations in specific frequency components in AM. The primary aim of this article is to identify the correct mode regardless of the control structure. To evaluate the capacity of ML algorithms to generalize classification to novel structures not present in the training data, a generalization test is conducted on a dataset of the viGFL structure, comprising 2500 samples. The viGFL structure differs only in the active power loop compared to the pqGFL structure. This modification induces a variation in the AM ranging from 1 Hz to 50 Hz on the dd axis.

Tab. 5 presents the results of generalization on the dataset using the viGFL structure. The percentage of generalization decreases compared to the classification. The decrease percentage varies among different ML algorithms, where some algorithms are less effective in generalization compared to others.
The analysis of the generalization results in this section is based on a comparison between two ML algorithms: DT and RF. It's easy to determine which columns or dimensions of the dataset these algorithms use for classification. Fig. 9 showcases the 5 most important features of both algorithms. In the labels, M or P represents the magnitude or phase, dd or qq represent the axes, and the number corresponds to the frequency (e.g., Pdd118Hz). An analysis of these important features reveals that in DT, one column holds more importance compared to other significant columns, whereas the distribution of importance in RF is almost equal among the columns. The capacity of RF to base its classification across multiple columns gives it a better generalization ability compared to a simple algorithm like DT, which relies on a single column. The developed ML algorithms were tested on a new dataset provided by Aalborg University, featuring structures outlined in the articles [11] [12]. Using 100 test samples for each GFM and GFL mode, the classification accuracy results are as follows: LR 92%, DT 83%, RF 95%, NBC 93%, XGB 87%, SVM 94%, KNN 94%.

Table 5: Generalization results

| ML algorithm | Generalization |
|---|---|
| LR | 85% |
| RF | 92% |
| DT | 40% |
| NBC | 82% |
| XGB | 79% |
| SVM | 84% |
| KNN | 84% |

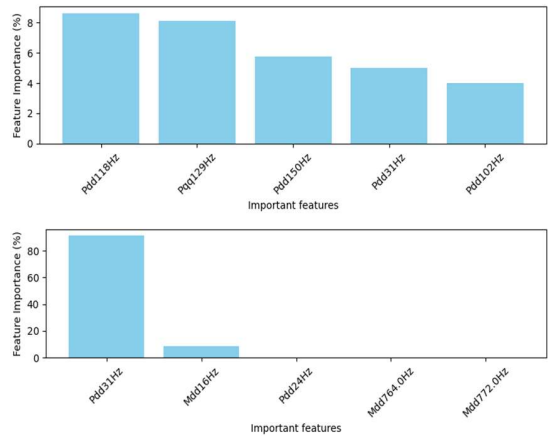

Figure 9: Important features of Random Forest and Decision Tree.

## 0.6 Conclusions

Results for the classification of operating modes of power converters from external measurements were presented in this paper. The frequency-domain admittance was studied throughout the validation of the external measurement method was investigated. Datasets are generated for converters operating in GFM or GFL modes, with two control structures for each mode using the external measurement method. Classical machine learning algorithms are developed with the aim of classifying these two control modes. The four control structures in the dataset have been successfully classified , but the ML algorithms fail to classify correctly an additional structure. The Random Forest algorithm proves to be most powerful in this application compared to others such as decision tree because its decision is based on the results of multiple trees, which are based on multiple dataset dimensions. For future works, new algorithms such as advanced neural networks are needed to allow the TSO's to classify correctly the operating, mode and control structure of power converters.

This work is part of the ANR DELTWINCO Project funded by the French National Research Agency (ANR) (LabEx BASC; ANR-21-CE05-0038)